\documentclass[conference]{IEEEtran}
\IEEEoverridecommandlockouts

\usepackage[utf8]{inputenc}
\usepackage[T1]{fontenc}
\usepackage{amsmath,amssymb,amsfonts}

\usepackage{bm}
\usepackage{mathtools}
\usepackage{graphicx}
\usepackage{url}
\usepackage{cite}
\usepackage{xcolor}
\usepackage{microtype}
\usepackage{balance}
\usepackage{amsthm}
\usepackage{booktabs}
\usepackage{enumitem}
\usepackage{tabularx}
\usepackage{soul}
\newtheorem{theorem}{Theorem}                % Theorem 1, 2, ...
\theoremstyle{definition}

\theoremstyle{remark}

\title{Dynamic Parameter Scheduling in\\ Soft-Hard BPGD for Lossy Source Coding}

\author{Masoumeh Alinia and David G. M. Mitchell \\
 Klipsch School
of Electrical and Computer Engineering, New Mexico State University, 
Las
Cruces, NM 88003\\ e-mail: \{aliniama,dgmm\}@nmsu.edu
\vspace{-4mm}
}
\date{April 2023}
\usepackage{amsfonts}
\usepackage{cite}

\DeclareMathOperator{\Ber}{Ber}

\begin{document}
\maketitle

\begin{abstract}
We investigate lossy source coding based on a soft-decision belief propagation guided decimation (BPGD) encoder for low-density generator matrix (LDGM) codes, referred to as \emph{soft-hard BPGD}. The performance of this encoder is highly sensitive to the choice of ``softness'' parameters, typically denoted by $(\beta,\mu)$, which are conventionally tuned via exhaustive empirical sweeps. % or phase-diagram heuristics. 
To reduce this burden and to better align the algorithm with the evolving graphical structure during decimation, we introduce a \emph{dynamic scheduling} framework in which $(\beta,\mu)$ are not fixed globally but change as decimation progresses. The schedule starts in a softer regime to encourage exploration and gradually hardens toward the end to promote convergence, similar to simulated annealing. We consider linear and exponential schedules, discuss their physical interpretation via an effective temperature viewpoint, and explain how they integrate with soft-hard BPGD without %significantly increasing its message update complexity.
changing the order of magnitude of its complexity.
Numerical experiments with irregular and semi-regular LDGM ensembles indicate improved rate-distortion performance and reduced non-convergence compared to constant-parameter baselines, while largely eliminating expensive grid searches for a single best pair $(\beta,\mu)$.
\end{abstract}
\section{Introduction}
Lossy source coding seeks to compress a source sequence into a compact representation that allows reconstruction within a prescribed distortion level \cite{Costa}. For binary symmetric sources under Hamming distortion \cite{Matsunaga}, the fundamental limit is given by Shannon's rate-distortion (RD) function. A variety of low-complexity encoders approach this limit, including trellis codes \cite{Viterbi}, polar codes\cite{Korada}, and sparse graph-based codes \cite{Wainwright}. In particular, LDGM codes \cite{wain}, the duals of LDPC codes, provide a sparse factorization of the reconstruction map and admit message-passing style algorithms.

%Plain belief propagation (BP) is not sufficient for lossy compression because multiple near-optimal reconstructions often coexist, leading to weakly biased or non-convergent marginals. Belief propagation guided decimation (BPGD) \cite{Filler}-\cite{alinia2} addresses this by iteratively computing biases, fixing a variable with the largest confidence, reducing the graph, and repeating on the smaller instance.
%The approach is strengthened by spatial coupling and by insights from spin glass theory \cite{aref} and \cite{alinia}, which connect algorithmic behavior to a Gibbs measure whose inverse temperature is proportional to a test-channel log-likelihood parameter.

Identifying encoding strategies for lossy source coding that are both simple to implement and capable of approaching the RD limit remains a central practical challenge. In this context, straightforward belief propagation (BP) is known to be inadequate: multiple near-optimal reconstructions typically coexist, producing weakly biased or non-convergent marginals and, consequently, poor encoder decisions. Belief propagation guided decimation (BPGD) addresses this limitation by iteratively computing biases, fixing one or more code nodes, reducing the graph, and repeating the procedure on the smaller instance. Variants include hard-decimation \cite{Filler}, soft-decimation \cite{castanheria}, and combined soft-hard BPGD \cite{alinia3, golmohammadi1}. %, with spatial coupling further strengthening performance by promoting wave-like progress along the chain and stabilizing reliable bias formation. 
From a statistical-physics perspective, these dynamics can be viewed through a Gibbs measure whose inverse temperature is proportional to a test-channel log-likelihood parameter, providing insight via spin-glass tools such as the cavity method \cite{aref}, \cite{alinia}.

Among practical encoders, soft decimation is attractive because it can achieve competitive distortion with linear complexity in the block length. Here, the hard decimation rule is replaced by a soft indicator (or reinforcement) that directs beliefs without committing prematurely. However, its performance depends critically on how the softness is parameterized. In particular, the choice of parameters that control generator node “temperature’’ and  code node reinforcement governs the trade-off between exploration (avoiding early freezing) and exploitation (locking into a consistent solution). To guide this choice, a cavity method–based framework has been proposed to compute critical thresholds for the softness parameters in soft and soft–hard BPGD \cite{alinia}. When parameters are selected close to these thresholds, soft–hard BPGD has been observed to outperform both hard- and soft-decimated variants in RD performance.

Despite these advances, finite-length effects still degrade performance. Cycles in the Tanner graph of the generator matrix induce message correlations during encoding, which can slow convergence and bias decisions away from globally good configurations. Moreover, conventional decimation policies either select nodes randomly \cite{Filler} or are based solely on instantaneous bias magnitude \cite{golmohammadi1}, without explicitly accounting for the local cycle structure that drives these correlations. To mitigate this, recent work \cite{alinia2} introduced a cycle-aware decimation strategy that periodically identifies code nodes participating in the densest short-cycle substructures and fixes them, thereby “breaking’’ problematic loops and reducing correlation in subsequent message updates.

In this paper, we combine these perspectives into a  unifying principle: both the graph and the effective algorithmic temperature should evolve during encoding. As the graph is decimated, degree distributions shift and local fields strengthen; correspondingly, the operating point that best balances exploration and exploitation should also change. Consequently, a single global choice for $(\beta,\mu)$ can be suboptimal across the entire trajectory. Motivated by this, the present work departs from fixed, globally tuned parameters and advocates \emph{dynamic} parameter scheduling within soft and soft–hard BPGD. In our approach, the encoder starts in a softer regime to avoid premature commitments and gradually hardens as a reliable structure emerges, which can be performed %, complementary to spatial coupling and compatible with cycle-aware policies, yielding improved distortion and convergence characteristics 
without significantly increasing per-iteration complexity. %In fact, to improve convergence stability and approach the optimal distortion–rate trade-off, the softness parameters $(\beta,\mu)$ are not kept fixed but are gradually adjusted during the decimation process. 
%The schedule begins in a soft regime corresponding to small $\xi$  (large $\beta$ and large  $\mu$), promoting exploration and preventing premature convergence. As $\xi$ increases through the schedule, both $\beta$ and $\mu$ decrease, gradually hardening the message updates and driving convergence toward a stable reconstruction.
We consider linear and exponential
schedules, discuss their physical interpretation via an effective
temperature viewpoint, and explain how they integrate with soft-hard BPGD without significantly increasing its message update
complexity. Numerical experiments with  irregular and semi-regular
LDGM ensembles indicate improved rate-distortion performance
and reduced non-convergence compared to constant-parameter
baselines, while largely eliminating expensive grid searches for a single best pair $(\beta, \mu)$.

%This paper is organized as follows. Section II introduces the LDGM code ensembles and notation used for lossy source coding. Section III reviews the soft–hard BPGD algorithm and explains its message-passing framework. Section IV presents the proposed dynamic parameter scheduling approach, describing both linear and exponential update strategies for the softness parameters. Section V analyzes the convergence of fixed-point iterations and discusses its implications for parameter scheduling in BPGD. Section VI reports numerical results comparing the scheduled and constant-parameter encoders in terms of distortion and convergence behavior. Finally, Section VII concludes the paper and outlines directions for future improvements and analytical extensions.

\section{LDGM Code Ensembles for Lossy Source Coding}\label{sec:background}
We consider a source $s=(s_1,\dots,s_N)$ with $s_i\overset{\text{i.i.d.}}{\sim}\Ber(1/2)$. LDGM codes, as duals of LDPC codes,  can be represented by a sparse generator matrix $\mathbf{G} \in\{0,1\}^{N \times M}$. We define the factor graph of this code as $\mathcal{G}=(V, C, E)$ where $V=\{1, \ldots, M\}$, $C= \{1, \ldots, N\}$, and $E=\{\ldots,(a, i), \ldots\}$ denote the sets of code bit nodes, the generator nodes and the edges connecting them, respectively. The vector  $(a,i)$ denotes an edge between generator node $a$ and code bit $i$, which occurs iff $\mathbf{G}_{a, i}=1$. We will use indices $a, b, c\in C$ to denote generator nodes and indices  $i, j, k \in V$ to denote code bits. We define the sets $C(i)=\{a \in C \mid(a, i) \in E\}$ and $V(a)=\{i \in V \mid(a, i) \in E\}$. { The coding (compression) rate is defined as $R \triangleq M/N$.} 
In a $(J,K)$-regular ensemble, all code nodes have degree $J$ and all generator nodes have degree $K$. In irregular ensembles, the connections between generator nodes and code bits in an optimized LDGM code are specified by a degree distribution $(\lambda, \rho)$ (from the edge perspective).

In this paper, we draw codes from both the Ising model construction of \cite{aref} and an optimized degree distribution \cite{castanheria}, which is optimized in a manner similar to density evolution in the analysis of LDPC codes for channel coding. In the Ising model, each edge emanating from a regular generator node with degree $K$ is connected uniformly at random to one of the code nodes. The degree of code nodes is a random variable with Binomial distribution $Bi(KN, 1/M)$, hence we refer to these codes as \emph{semi-regular}. In the asymptotic regime of large $N$, $M$, the code node degrees have i.i.d. Poisson distribution with an average degree $K/R$. 
%The connections  between generator nodes and
%code bits in an optimized LDGM code are specified by
%a degree distribution $(\lambda,\rho)$ from the edge perspective,
%$\rho(x)=\sum_{i}\rho_{i}x^{i-1}$ and $\lambda(x)=\sum_{i}\lambda_{i}x^{i-1}$, where $\rho _i$
%and $\lambda _{i}$ denote the proportion of all edges connected to generator
%nodes and bit nodes with degree $i$, respectively.
%We consider both regular and irregular LDGM ensembles. In a $(J,K)$-regular ensemble, all check nodes have degree $J$ and code bit nodes have degree $K$. In irregular ensembles, the connections between generator nodes and code bits in an optimized LDGM code are specified by a degree distribution $(\lambda, \rho)$ (from the edge perspective) as
%$$\rho(x) = \sum_i \rho_i x^{i-1}, \qquad
%\lambda(x) = \sum_i \lambda_i x^{i-1},$$
%where $\rho_i$ and $\lambda_i$ denote the proportion of all edges connected to generator nodes and bit nodes with degree $i$, respectively.
 
 Given a codeword $w\in\{0,1\}^{M}$, the reconstruction is $\hat{\mathbf{s}}=\mathbf{G}\mathbf{w}$ over $\mathbb{F}_2$. Distortion is measured by the relative Hamming distance
\begin{equation}
d(s,\hat{s})=\frac{1}{N}\sum_{i=1}^{N}|s_i-\hat{s}_i|.
\end{equation}
Shannon's RD function for the Bernoulli $(1/2)$ source is $R(D)=1-h_2(D)$ where $h_2(\cdot)$ is the binary entropy function. We interpret encoding as inference under a Gibbs posterior
\begin{equation}
P(w\mid s) \propto \exp\!\big(-\gamma \, d\big(s,\mathbf{G}\mathbf{w}\big)\big),
\end{equation}
where $\gamma>0$ plays the role of an inverse temperature. In this view, maximizing $P$ corresponds to minimizing distortion, and belief propagation computes approximate marginals. Because the posterior typically has many near-minimizers, decimation is used to progressively reduce the solution space.

\section{Soft-Hard BPGD}
Soft-hard BPGD combines soft reinforcement \cite{castanheria} inside the BP updates with (optional) hard code node fixing between iterations \cite{Filler}. Let 

\begin{equation}\label{eq1}
R_{i}^{(t+1)}=\sum_{a \in C(i)} \widehat{R}_{a \rightarrow i}^{(t)}, \quad R_{i \rightarrow a}^{(t+1)}=\sum_{b \in C(i) \backslash a} \widehat{R}_{b \rightarrow i}^{(t)}+\frac{1}{\mu} R_{i }^{(t)},
\end{equation}
\begin{equation}\label{eq2}
\widehat{R}_{a \rightarrow i}^{(t+1)}=2(-1)^{s_{a}+1} \tanh ^{-1}\left(\beta \prod_{j \in V(a) \backslash i} B_{j \rightarrow a}^{(t)}\right),
\end{equation}
\begin{equation}\label{eq3}
B_{i}^{(t)}=-\tanh \left(\frac{R_{i}^{(t)}}{2}\right), \quad B_{i \rightarrow a}^{(t)}=-\tanh \left(\frac{R_{i \rightarrow a}^{(t)}}{2}\right),
\end{equation}
where $R_{i \rightarrow a}^{(t)}$, $\widehat{R}_{a \rightarrow i}^{(t)}$, and $B_{i \rightarrow a}^{(t)}$ denote, respectively, the message from code node $i$ to generator node $a$, the message from generator node $a$ to code node $i$, and the bias associated with $R_{i \rightarrow a}^{(t)}$ at iteration $t$. Likewise, $R_{i}^{(t)}$ and $B_{i}^{(t)}$ are the LLR of code bit $i$ and its associated bias. The quantities $\beta=\tanh(\gamma)$ and $\mu$ are nonnegative parameters. The parameter $\gamma$ controls how aggressively the message passing procedure attempts to make the reconstruction $\hat{\mathbf{s}}=\mathbf{G}\mathbf{w}$ align with the source $\mathbf{s}$, a larger $\gamma$ enforces a stronger drive, although the code structure ultimately limits this. 

The additive term $\frac{1}{\mu}R_{i}^{(t)}$ in \eqref{eq1} serves as a soft reinforcement. %, yielding a relaxed counterpart of hard decimation within the BP updates.
The soft indicator
\[
I_S\!\big(B_{i \rightarrow a}^{(t)}\big)=\frac{2}{\mu}\tanh^{-1}\!\big(B_{i \rightarrow a}^{(t)}\big)=\frac{1}{\mu}R_{i \rightarrow a}^{(t)}
\]
approximates the hard–indicator function~\cite{golmohammadi1},
\[
I_H\!\left(B_{i \rightarrow a}^{(t)}\right)=
\begin{cases}
-\infty, & B_{i \rightarrow a}^{(t)}=-1,\\
0, & -1<B_{i \rightarrow a}^{(t)}<1,\\
+\infty, & B_{i \rightarrow a}^{(t)}=1,
\end{cases}
\]
with $\mu$ governing the degree of softness and therefore referred to as the \emph{softness parameter}. In this framework, $\beta$ controls the effective inverse temperature at generator nodes, determining how strongly parity consistency is enforced, while $\mu$  regulates the strength of code node reinforcement. Large $\mu$ weakens reinforcement, producing softer updates, whereas small $\mu$ leads to harder, more decisive updates. Thus, the overall softness of the algorithm depends jointly on $(\beta, \mu)$.
%Algorithm~\ref{alg:cap} outlines the procedure, where $t$ indexes the iteration, $\mathcal{G}^{(t)}$ is the LDGM graph at iteration $t$, and $w_i$ is the binary assignment made to code node $i$.
The initial code to generator messages $R_{i \rightarrow a}^{(0)}$ are drawn from $\{\pm 0.1\}$ with $\mathbb{P}\!\left(R_{i \rightarrow a}^{(0)}=0.1\right)=0.5$, and they are reset to $0$ at iteration~1. After several message
 updates, the algorithm fixes the most biased bit (breaking
 ties randomly), removes it and its incident edges, updates the
 affected generators, and repeats on the reduced graph. Stopping
 can occur when all bits are fixed or when a maximum number
 of decimation rounds is reached. This procedure retains the
 linear per–iteration complexity characteristic of sparse message
 passing.

There is no closed–form rule to choose $\mu$ or $\beta$ that guarantees optimal distortion; instead, \cite{castanheria} and subsequent works (e.g., \cite{golmohammadi1}) determine a good $\mu$ through exhaustive, computationally intensive simulations. % with pure grid approach. Pure grid approach refers to a parameter-sweep simulation method, testing the algorithm over a fixed grid of possible parameter values to find the best performing pair $(\beta,\mu)$. 
In \cite{alinia}, the %leverages spin-glass insights via the 
cavity method is applied to calibrate the softness parameters for the soft BPGD algorithm, yielding parameter choices that provide strong RD performance. All these approaches are \emph{constant parameter implementations}, i.e., $\beta$ and $\mu$ are set once and remain fixed for the entire algorithm. While effective when optimally tuned, such fixed settings can be mismatched to early or late stages of decimation, leading respectively to premature freezing or sluggish convergence.  
 In the next section, we apply a scheduled dynamic selection of parameter to combat this problem 
in the soft and 
soft-hard BPGD algorithm to ensure good RD performance. {The term ``scheduling'' here refers to varying $(\beta,\mu)$ across decimation rounds %
%to induce a soft-to-hard annealing behavior, 
and should not be confused with the message-passing
update order used in belief propagation.}

\section{Dynamic Parameter Scheduling in Soft and Soft--Hard BPGD}

The main limitation of a grid search approach is that the computational complexity is high and, as such, conventional approaches employ a single fixed pair of parameters $(\beta,\mu)$ across all rounds of decimation; however, empirical studies show that this is not optimal (see Section \ref{sec:results}).  %During the early rounds of decoding, using softer values of $(\beta,\mu)$ helps prevent overconfident but incorrect decimations. In contrast, during the later rounds, stronger values of $\beta$ and $\mu$ are required to drive convergence. As a result, no single fixed parameter pair is globally optimal. Better rate--distortion performance can often be achieved by \emph{adapting} $(\beta,\mu)$ during decoding through a linear or exponential scheduling strategy.
The central idea of this work is to \emph{schedule} $(\beta_r,\mu_r)$ as the decimation (or iteration) round $r$ progresses. Let the decimation index be $r=0,1,\dots,\nu-1$, where $\nu$ is the total number of  rounds. In this work, we select a  pair $(\xi_{\text{start}},\xi_{\text{end}})$  with $\xi_{\text{start}}<\xi_{\text{end}}<1$ and determine the parameters for round~$r$ using an auxiliary variable~$\xi$ according to
\begin{equation}
\beta_r = \frac{1-\xi_r}{1+\xi_r}, \qquad
\mu_r = \frac{1}{\xi_ r}.
\label{eq:xi_relation}
\end{equation}
With a monotone increase $\xi_{\text{start}}<\xi_r<\xi_{\text{end}}$, we have that $\beta_r$ and $\mu_r$ are both decreasing
functions of $r$, while the reinforcement weight $1/\mu_r=\xi_r$ increases with $r$.

Intuitively, decreasing $\beta_r$ reduces the multiplicative gain in the generator-node nonlinearity (and can mitigate
oscillations when intermediate messages become large), whereas decreasing $\mu_r$ increases the effective reinforcement
term through $1/\mu_r$, strengthening the additive bias injected into the code-node update.
Consequently, although $\beta_r$ becomes ``softer'' in the sense of reducing generator gain, the increasing reinforcement
$1/\mu_r$ drives the overall behavior toward a more decisive (harder) regime as decimation progresses.
This combined evolution is particularly useful near the decimation threshold, where overly large generator gain can
destabilize iterations, yet stronger reinforcement is needed to lock in reliable variable assignments.

 %During the iterative decoding process, the parameters $\beta_r$ and $\mu_r$
 %are gradually varied to balance exploration and convergence.
%At the early stages,  a lower value of $\xi$ (larger values of $\beta<1$) and larger $\mu$ keep the message updates soft, preventing premature saturation of log-likelihood ratios (LLRs) and allowing information to propagate more evenly through the factor graph.
%As iterations proceed and local agreement among code nodes increases, both parameters are progressively decreased, where $\beta$  and 
%$\mu$ takes smaller values that enforce the constraints more strongly and drives the system toward a stable, hard decision.
%This soft-to-hard transition helps avoid trapping in incorrect fixed points while ensuring final convergence to a valid reconstruction near the end of the decoding process.

Two forms of scheduling are considered: \emph{linear}, and \emph{exponential (geometric)}. Here, the evolution of~$\xi$ over decimation round $r$ is given as:
\begin{enumerate}[label=(\alph*)]
    \item {\bf Linear schedule:}
The progression fraction is $t_r = r/(\nu-1)$, giving
\begin{equation}
\xi_r = \xi_{\text{start}} + t_r\big(\xi_{\text{end}}-\xi_{\text{start}}\big);
\label{eq:xi_linear}
\end{equation}
\item {\bf Exponential (geometric) schedule:}
\begin{equation}
\xi_r = \xi_{\text{start}}
\left( \frac{\xi_{\text{end}}}{\xi_{\text{start}}} \right)^{t_r}.
\label{eq:xi_exponential}
\end{equation}
\end{enumerate}
We note here that in the case of soft-hard or hard decimation, the decimation round $r$ can have $t\geq 1$ rounds of algorithmic iteration. In the case of the {soft BPGD} variant, the same scheduling law applies, but the parameters evolve continuously with the iteration index $t$ rather than discrete decimation steps, i.e., $r=t$.

The {linear schedule}, given by~(\ref{eq:xi_linear}), provides a simple uniform progression of the softness parameters across rounds, ensuring a steady and predictable transition from soft to hard behavior. This linear evolution is often sufficient when message reliabilities increase roughly uniformly during decimation, offering an interpretable baseline against which more adaptive schedules can be compared. The exponential schedule \eqref{eq:xi_exponential} ensures a constant multiplicative change in~$\xi$ across rounds, equivalent to equal steps in $\log\xi$.  
This yields a smooth, scale-invariant ``cooling'' process that avoids abrupt transitions and has proven effective in stabilizing belief-propagation dynamics~\cite{mooij-kappen-uai}.%, especially in large spatially coupled graphs. From an annealing viewpoint, early iterations benefit from a smaller ``temperature'' (higher~$\beta$, moderate~$\mu$) to avoid poor local minima, while later iterations require a warmer regime to consolidate parity consistency.  
%Spatially coupled constructions tolerate larger $\beta_{\text{end}}$ due to wave-like bias propagation, while irregular degree profiles may benefit from larger $\mu_{\text{start}}$ to stabilize early updates.   
{ Empirically, exponential schedules tend to be more robust for denser constraints (e.g., larger generator degree $K$,
more short cycles, and longer blocks), since they harden gently in early rounds and accelerate hardening near the end.
Linear schedules are often sufficient for sparser graphs and smaller $K$, where bias reliability improves more uniformly.
These trends are corroborated by the numerical results in Section }\ref{sec:results}.

\section{Convergence of Fixed-Point Iterations and Implications for Parameter Scheduling in BPGD}
\label{sec:beta-schedule-convergence}
This section discusses our motivation and justifies a monotone (soft-to-hard) scheduling of the parameters $\beta$ and $\mu$ within the message updates used for soft decimation equations (\ref{eq1})–(\ref{eq3}). Our justification is grounded in standard fixed-point convergence theory and known sufficient conditions for convergence of loopy BP. 

%BP on loopy graphs can be viewed as a deterministic map on the vector of messages; it seeks a fixed point of that map.
BP on loopy graphs can be viewed as a deterministic self-map on the vector of messages, i.e., a mapping $F:\mathbb{R}^{|E|} \rightarrow \mathbb{R}^{|E|}$ that transforms the current set of messages into their updated values. The fixed points of this map correspond to consistent beliefs under the approximate Gibbs measure.
Convergence properties of this iterative map have been analyzed through sufficient conditions ensuring the existence and uniqueness of fixed points, as well as contractivity under suitable norms. Two complementary lines of results can be identified in \cite{ortega-rheinboldt,kelley,peters-fixedpoint}: (i)~those establishing sufficient conditions for a unique fixed point and guaranteed convergence, typically expressed via norm or spectral-radius bounds on the update operator, and (ii)~those relying on perturbation or error-accumulation analyses that likewise imply convergence. Across these works, a common principle emerges: the effective local \emph{gain} of the BP update \cite{mooij-kappen-uai,ihler-tr} must remain below unity, ensuring that the Jacobian’s spectral radius stays smaller than one in a neighborhood of the desired fixed point, thereby preserving contraction and stability.

{Stability primarily ensures that the BP iterations do not diverge or oscillate, so the message updates remain numerically well-behaved throughout the encoding process. Achieving low distortion, however, requires the algorithm to operate in parameter regimes where the induced Gibbs measure concentrates around high-quality configurations that are near the optimal rate–distortion solution. In practice, the proposed scheduling improves the achieved rate–distortion performance empirically by keeping BP in a stable and informative regime during early decimation rounds and progressively strengthening the bias as the algorithm transitions toward hard decisions.}

\subsection{Fixed-point Iterations: Contraction and Local Linearization}

%Let $x^{(t)}$ denote the vector state of the solution at time $t$ so that $x^{(t+1)} = F(x^{(t)})$ corresponds to an iteration where a fixed point $x^\star$ exists (i.e., $F(x^\star)=x^\star$).Two classical facts explain when nearby iterates (an iteration of a state) converge to $x^\star$.
Let $x^{(t)}$ denote the \emph{state vector} of BP at iteration~$t$, and let $x^{(t+1)} = F(x^{(t)})$ represent one full update of the BP map $F$. A fixed point $x^\star$ satisfies $F(x^\star) = x^\star$, corresponding to a self-consistent set of messages. We now state two  results that characterize when \emph{nearby state vectors} $x^{(t)}$ in the message space converge to this fixed point $x^\star$.
\begin{theorem}[Local convergence via Jacobian spectral radius~\cite{ortega-rheinboldt}]
\label{thm:local}
Assume $F$ is continuously differentiable near a fixed point $x^\star$.
%If the Jacobian $J=\left.\frac{\partial F}{\partial %x}\right|_{x^\star}$ satisfies
%$\rho(J)<1$ (spectral radius), then $x^\star$ is a \emph{locally %attracting} fixed point.
If the Jacobian $J=\left.\frac{\partial F}{\partial x}\right|_{x^\star}$ satisfies
$\rho(J)<1$\footnote{The spectral radius $\rho(J)$ is defined as
$\rho(J)=\max_i |\lambda_i(J)|$, where $\lambda_i(J)$ are the eigenvalues of $J$ \cite{mooij-kappen-uai}.},
then $x^\star$ is a \emph{locally attracting} fixed point.
Then
there exists a neighborhood $\mathcal{U}$ of $x^\star$ such that for any $x^{(0)}\in\mathcal{U}$,
the iterates are well-defined and $x^{(t)}\to x^\star$ with asymptotic rate $\|x^{(t+1)}-x^\star\|
\approx \rho(J)\,\|x^{(t)}-x^\star\|$.
\end{theorem}
%\noindent
Intuitively, $\rho(J)$ is the iteration's local ``gain''. If all linearized modes are shrunk ($\rho(J)<1$), errors decay; if some mode is amplified ($\rho(J)>1$), errors grow (oscillations/divergence)   \cite{ortega-rheinboldt, kelley}.
A global guarantee is provided by the Banach contraction theorem: if $F$ is a contraction with constant $q\in[0,1)$ on a complete metric space, then a unique fixed point exists and Picard iteration converges to it from any initial point.
\begin{theorem}[Global (Banach) contraction \cite{peters-fixedpoint}]
If $F$ is a contraction on a complete metric space,
$\|F(x)-F(y)\|\le q\|x-y\|$ for some $q\in[0,1)$, then a unique fixed point exists and
$x^{(t)}\to x^\star$ for any start $x^{(0)}$;
moreover $\|x^{(t)}-x^\star\|\le \frac{q^t}{1-q}\|x^{(1)}-x^{(0)}\|$.
\end{theorem}

% &&&&&&&&&&&&&&&&&&&&&&&&&&&&&&&&&&&&&&&&&&&&&&&&&&&&&&&&&&&&&&&&&&&&&&&&&&&&&&&&&&&&&&&
\subsection{Relationship Between the BP Update, $\beta$, Perturbations, and Contraction}
Let $m^{(t)}$ denote the BP state  at iteration $t$. One full ``BP sweep'' defines a $\beta$–parametrized map (iteration)
\begin{equation}
  m^{(t+1)} \;=\; F_{\beta}\!\bigl(m^{(t)}\bigr).
\end{equation}
Let $m^\star$ be a fixed point, $F_{\beta}(m^\star)=m^\star$, and define the perturbation
$\delta^{(t)} = m^{(t)}-m^\star$.
Linearizing at $m^\star$ yields
\begin{equation}\label{eq10}
  \delta^{(t+1)} \;\approx\; J_{\beta}\,\delta^{(t)},
  \qquad
  J_{\beta} \;=\; \left.\frac{\partial F_{\beta}}{\partial m}\right|_{m^\star}.
\end{equation}
Hence, for any operator norm $\|\cdot\|$,
\begin{equation}
  \|\delta^{(t+1)}\| \;\le\; \|J_{\beta}\|\,\|\delta^{(t)}\| .
\end{equation}
If there exists $q<1$ with $\|J_{\beta}\|\le q$, then
$\|\delta^{(t)}\| \le q^{\,t}\|\delta^{(0)}\|$, i.e., the update is locally \emph{contractive} \cite{mooij-kappen-uai}.
Equivalently, local convergence holds if the spectral radius satisfies $\rho(J_{\beta})<1$.

{Equation}~\eqref{eq10} {follows from a first-order Taylor expansion of the BP update map}
$F_{\beta}(\cdot)$ around a fixed point $m^{\star}$.
Specifically, for $m = m^{\star} + \delta$, we have
\begin{equation}
F_{\beta}(m) \;=\; F_{\beta}(m^{\star}) + J_{\beta}\,\delta + \mathcal{O}\!\left(\|\delta\|^{2}\right),
\end{equation}
%where $J_{\beta}$ denotes the Jacobian of $F_{\beta}$ evaluated at $m^{\star}$.
%Since $F_{\beta}(m^{\star}) = m^{\star}$, it follows that the perturbation evolves as
%\begin{equation}
%\delta^{(t+1)} \;=\; J_{\beta}\,\delta^{(t)} + \mathcal{O}\!\left(\|\delta^{(t)}\|^{2}\right),
%\end{equation}
%which explains the use of the approximation symbol in \eqref{eq10}.
The relation becomes exact when the higher-order terms vanish (e.g., $F_{\beta}$ is locally affine),
and is {asymptotically tight in the limit $\|\delta^{(t)}\|\to 0$.}

{\noindent\textit{Remark.}
Since decimation alters the graph structure between rounds, the stability and contraction
conditions derived in this section characterize BP behavior on a fixed reduced graph.
They therefore provide heuristic insight for guiding parameter schedules, rather than
formal global guarantees for convergence or rate--distortion optimality of the full BPGD encoder.}
\subsection{Influence of the Inverse Temperature Parameter $\beta$.}
The parameter $\beta$ contributes to the BP iteration through the generator-to-code update \eqref{eq2}. Each outgoing message from a generator node $a$ to a code node $i$ at iteration $t{+}1$ is in the form of
$$\widehat{R}^{\,t+1}_{a\to i}
  = 2(-1)^{s_a+1}\tanh ^{-1}\!\big(u_{a\to i}\big),
 $$
 with
  \begin{equation}\label{state}u_{a\to i} = \beta \prod_{j\in V(a)\setminus i}\! B^{\,t}_{j\to a}.
\end{equation}
Hence, $\beta$ acts as a global gain factor inside the nonlinear mapping $\tanh ^{-1}(\cdot)$. Differentiating this update with respect to one neighboring message $ B^{\,t}_{k\to a}$ gives the corresponding local sensitivity,
\[\displaystyle
\frac{\partial \widehat{R}^{\,t+1}_{a\to i}}{\partial  B^{\,t}_{k\to a}}
  = 2(-1)^{s_a+1}
   \frac{1}{1-u_{a\to i}^{2}}
   \Bigl(\beta\!\!\prod_{j\neq i} B^{\,t}_{j\to a}\Bigr)
   \frac{1}{ B^{\,t}_{k\to a}},
\]
whose magnitude can be expressed as
\[
\Bigl|\tfrac{\partial \widehat{R}^{\,t+1}_{a\to i}}
              {\partial  B^{\,t}_{k\to a}}\Bigr|
 = \frac{2\,|u_{a\to i}|}{|1-u_{a\to i}^{2}|}
   \frac{1}{| B^{\,t}_{k\to a}|},
\]
%This expression reveals that $\beta$ appears linearly in $u_{a\to i}$ while also influencing the denominator $1-u_{a\to i}^{2}$. 
As $\beta$ increases, the term $|u_{a\to i}|$ grows linearly, whereas the denominator diminishes, leading to an overall super-linear amplification of the Jacobian entries. The subsequent code node update,
\[
 R^{\,t+1}_{i}=\sum_{b\in C(i)}\widehat{R}^{\,t}_{b\to i},
\qquad
 B^{\,t+1}_{i}=-\tanh\!\Big(\frac{ R^{\,t+1}_{i}}{2}\Big),
\]
introduces an additional attenuation factor
\[
\frac{\partial {B}^{\,t+1}_{i}}{\partial {R}^{\,t+1}_{i}}
  = -\tfrac{1}{2}\,\operatorname{sech}^{2}\!\Big(\tfrac{{R}^{\,t+1}_{i}}{2}\Big)
  \in (0,\tfrac{1}{2}].
\]
which moderates but does not eliminate the strong dependence on $\beta$. 
%When these layers are combined, each effective Jacobian element scales approximately as
Multiplying the local derivatives of the generator and code-node update equations  obtains the effective Jacobian gain approximation for a full BP iteration  as
\[
\text{gain} \;\propto\;
\frac{|u_{a\to i}|}{1-u_{a\to i}^{2}}
\times (\text{degree factors})
\times
\operatorname{sech}^{2}\!\Big(\tfrac{ R}{2}\Big).
\]

Consequently, increasing $\beta$ monotonically enlarges the magnitudes of the Jacobian entries, thereby increasing the spectral radius $\rho(J_\beta)$. Since $\beta$ pushes $|u_{a\to i}|$ toward unity, the denominator $1-u_{a\to i}^{2}$ approaches zero and the spectral radius can exceed one, violating the contraction condition and producing oscillatory or divergent behavior. In contrast, very small $\beta$ values suppress the message gain, leading to overly damped iterations that converge slowly to suboptimal fixed points associated with higher distortion.

Hence $\beta$ functions as a global temperature-like parameter that continuously adjusts the effective local gain of the BP dynamics. An intermediate value $\beta^*$ typically satisfies $\rho(J_\beta)\!\lesssim\!1$, preserving stability while retaining sufficient nonlinearity to achieve low distortion. In practice, we have observed that reliable performance is obtained by beginning with a relatively large value of $\beta$ (close to~1) and then gradually decreasing it, thereby keeping $|u_{a\to i}|=\beta\prod| B_{j\to a}|$ safely below unity while exploring the minimum-distortion regime.
%Observed oscillations indicate that $\rho(J_\beta)\!>\!1$ and that $\beta$ should be reduced, whereas excessively slow convergence or plateaued distortion suggest increasing $\beta$ slightly. Through this mechanism, $\beta$ directly governs both the contraction property of the fixed-point map and the attainable distortion in soft-decimation BP.

Although both $\beta$ and $\mu$ influence the convergence of the BP fixed-point map, we emphasize $\beta$ because it most strongly controls the effective local gain at generators and thus the attainable distortion. The parameter $\mu$, which enters the code node to the generator update equation (\ref{eq1}) as a damping factor $1/\mu$, primarily governs stability and smoothness of the iteration: increasing $\mu$ attenuates message updates and promotes contraction, while smaller $\mu$ sharpens updates and can accelerate convergence but risks oscillations. Accordingly, we do not neglect $\mu$; rather, we hold it in a stabilizing range and focus our analysis on $\beta$, whose modulation of the generator nonlinearity has the dominant impact on distortion performance. In our schedule, the auxiliary variable $\xi$ increases over rounds, causing $\beta$ to decrease while $1/\mu$ increases (annealing in the code node layer), strengthening the additive reinforcement. Although $\beta$ softens the multiplicative gain, the growing reinforcement $1/\mu$ dominates near the decimation threshold, producing an overall soft-to-hard evolution of the algorithm.

%&&&&&&&&&&&&&&&&&&&
\subsection{Verification of the Theorem~1 Conditions}
We consider the BP update as a mapping on  $F_\beta:\mathcal{D}_\varepsilon\rightarrow\mathcal{D}_\varepsilon$ 
defined on the \emph{clipped message domain}
\[
\mathcal{D}_\varepsilon = (-1{+}\varepsilon,\,1{-}\varepsilon)^{|E|},
\]
where $E$ is the set of undirected edges 
in the factor graph, $\varepsilon>0$ is a small numerical margin (e.g., $10^{-6}$) used to keep all message values
away from the singular points $\pm1$ and the state is $m^{(t)} = \{u_{a\to i}\}\in \mathcal{D}_\varepsilon$.
This clipping ensures that the nonlinear functions $\tanh(\cdot)$ and $\operatorname{atanh}(\cdot)$
used in the code node and generator updates remain bounded and continuously differentiable.
Consequently, the Jacobian $J_\beta(x)$ exists and is continuous for all $x\in\mathcal{D}_\varepsilon$.

The local derivatives of the generator and code node messages yield the following for any edge pair $(k\!\to\!a,\,a\!\to\!i)$:
\begin{align*}
   \left|\frac{\partial \widehat{R}_{a \to i}}{\partial B_{k \to a}}\right|
 &= \frac{2\,|u_{a \to i}|}{|1 - u_{a \to i}^2|} \cdot \frac{1}{|B_{k \to a}|},\\ 
u_{a \to i}  &= \beta \prod_{j \in V(a)\setminus i} B_{j \to a},\\
\left|\frac{\partial B_i}{\partial R_i}\right|  &= \tfrac{1}{2}\,\text{sech}^2(R_i/2) \le \tfrac{1}{2}.
\end{align*}
\noindent With clipping/floors on $\mathcal{D}_\varepsilon$, we have 
$|B_{j \to a}| \le 1-\varepsilon$,
%$|B_{k \to a}| \ge \varepsilon$, 
hence
\[
|u_{a \to i}| \le \beta(1-\varepsilon)^{d_c-1}, 
\quad |1 - u_{a \to i}^2| \ge 1 - \beta^2(1-\varepsilon)^{2(d_c-1)}.
\]
\noindent Therefore, each one-hop entry (incoming $B_{k \to a}$ to outgoing $B_i$) is bounded by
\begin{align}
\left|\frac{\partial B_i}{\partial B_{k \to a}}\right|
&\le \underbrace{\tfrac{1}{2}}_{\text{code node}} 
  \underbrace{\frac{2\beta(1-\varepsilon)^{d_c-2}}
  {1 - \beta^2(1-\varepsilon)^{2(d_c-1)}}}_{\text{generator}}  \nonumber\\
  %\underbrace{\frac{1}{\varepsilon}}_{\text{floor}}
&= \frac{\beta(1-\varepsilon)^{d_c-2}}
{[1 - \beta^2(1-\varepsilon)^{2(d_c-1)}]}.
\end{align}
%\varepsilon
We now sum the contributions for all neighbors that feed code node $i$ in one round.
A code node of degree $d_v$ receives at most $d_v$ generator messages, and each of those depends on up to $(d_c - 1)$ incoming biases. 
A safe row-sum bound is thus
\begin{align}
\|\mathbf{J}_\beta(x)\|_\infty 
&\le (d_v)(d_c - 1)\,
\frac{\beta(1-\varepsilon)^{d_c-2}}
{[1 - \beta^2(1-\varepsilon)^{2(d_c-1)}]} \nonumber\\
&=:\, L(\beta,\varepsilon,d_v,d_c).\label{eq:Lbeta}
\end{align}
%\varepsilon
Here “safe row-sum bound’’ means a conservative upper bound on $\|J_\beta(x)\|_\infty$ obtained by summing worst-case absolute one–hop sensitivities across all entries in a row, yielding \eqref{eq:Lbeta} and guaranteeing $\rho(J_\beta)\le \|J_\beta\|_\infty$.
Choosing $\beta$ and $\varepsilon$ such that $L(\beta,\varepsilon,d_v,d_c)<1$
guarantees $\|J_\beta(x)\|_\infty<1$ for all $x\in\mathcal{D}_\varepsilon$,
and therefore $\rho(J_\beta)\le\|J_\beta\|_\infty<1$.
Under these conditions, the BP update map $F_\beta$ is a contraction on $\mathcal{D}_\varepsilon$.
Hence, the hypotheses of Theorem~1 (existence, differentiability, and contractivity of the update map)
are satisfied for the range of $\beta$ values used in the soft regime.
This ensures the existence and uniqueness of a fixed point and explains the stable convergence
observed in our numerical results. From the contraction analysis, smaller $\beta$ values ensure the spectral radius of the BP Jacobian remains below one, maintaining stable convergence. In practice, the algorithm begins with a higher $\beta$ (close to 1), exploratory updates, and gradually decreases $\beta$ to sustain contraction while enforcing stronger parity consistency near convergence.
%\hl{We stress that} \eqref{eq:Lbeta}\hl{ provides a conservative sufficient condition for local stability.
%In practice, BP iterations often remain stable for larger values of $\beta$ than those predicted
%by the bound. Moreover, the parameter $\epsilon$ reflects numerical clipping used to prevent
%message saturation, rather than a fundamental property of the underlying model.}
\section{Numerical Results}\label{sec:results}
 In this section, the results of various experiments of a Python
based implementation of BPGD encoders are reported for different
 LDGM code ensembles. As described in Section \ref{sec:background}, codes are generated randomly
 following the Ising model or the optimized degree distribution
 from  \cite{castanheria},
%We evaluate the proposed schedules on regular $(J,K)$ spatially coupled LDGM codes with coupling width $w$ and length $L$ and  an irregular LDGM ensemble characterized by the edge-perspective degree distributions
\begin{align}
\lambda(x) &= x^{6}, \nonumber\\
\rho(x) &= 0.275698\,x + 0.25537\,x^{2} 
         + 0.076598\,x^{3} + 0.39233\,x^{8}.
\end{align}
 For all experiments, the source is Bernoulli-$(1/2)$, the reconstruction is $\hat{\mathbf{s}}=\mathbf{G}\mathbf{w}$ over $\mathbb{F}_2$, and distortion is measured by the normalized Hamming distance. We compare three soft and soft-hard BPGD encoders: a constant parameter  baseline tuned by a modest grid over $(\beta,\mu)$, a linear scheduled encoder and exponentially scheduled encoder, where the schedules follow  $(\xi_{\text{start}},\xi_{\text{end}})$. In all cases, we enforce the same message-update limits per round and the same decimation budget. The values of $\xi_{\text{start}}$ and $\xi_{\text{end}}$ were determined through short exploratory runs 
to balance early‐stage stability and late‐stage convergence. 
A small $\xi_{\text{start}}$ yields a larger $\beta$ and $\mu$ (softer updates) that prevents premature freezing, 
whereas a larger $\xi_{\text{end}}$ reduces $\beta$ and $\mu$, hardening the decisions 
as reliable biases emerge. 
For all reported results, $\xi_{\text{start}}$ and $\xi_{\text{end}}$ were chosen 
from a narrow neighborhood around the empirically optimal constant‐$\xi$ baseline.

\begin{figure}[t]
	\begin{center}
		\includegraphics[width=\columnwidth]{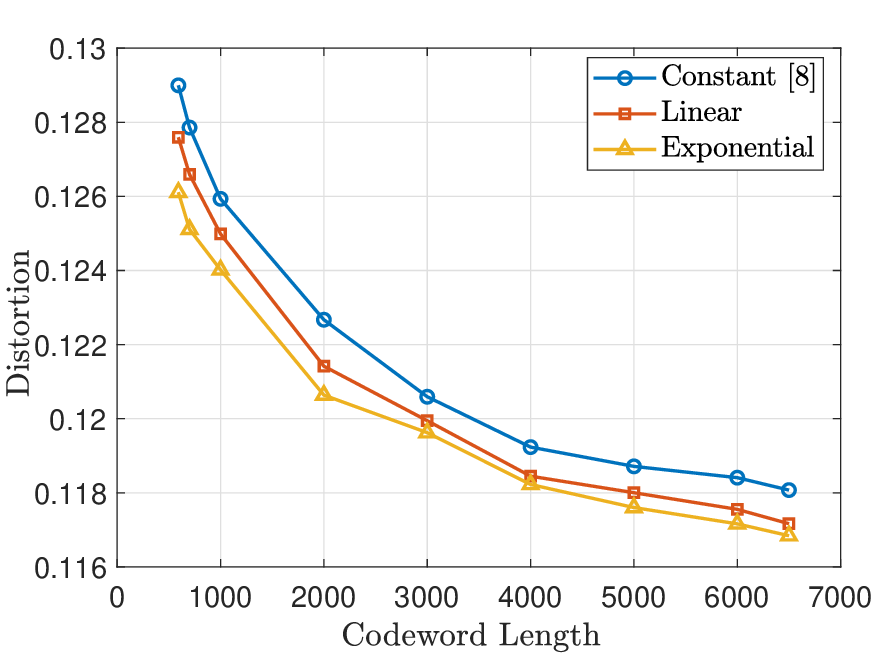} \end{center}\vspace{-5mm}
	\caption{Distortion performance over a range of  code lengths with  soft BPGD.}\label{fig1}%\vspace{-3mm}
\end{figure}

Figure \ref{fig1} shows the average distortion obtained using soft-BPGD encoding for the optimized irregular LDGM ensembles \cite{castanheria} under the three different scheduling strategies for the softness parameters $(\beta,\mu)$. %Here, the  constant baseline tuned following \cite{castanheria}, and the proposed linear and exponential schedules. 
%All simulations use irregular LDGM ensembles \cite{castanheria} with identical degree distributions and identical iteration budgets. 
As the codeword length increases, all methods approach lower distortion due to concentration effects, but both of the scheduled schemes are observed to consistently outperform the constant baseline. The exponential schedule achieves the lowest distortion across nearly all lengths, reflecting its smoother “cooling” behavior that preserves stability in early iterations while gradually enforcing stronger parity consistency. The linear schedule also yields clear gains for moderate block lengths, indicating that dynamically adapting the softness parameters improves convergence and rate-distortion performance. %Figure \ref{fig1} illustrates results for the soft BPGD variant, while 
%Table \ref{tab1} reports the corresponding results for the soft–hard BPGD encoder, where additional hard variable fixing is applied between message-passing rounds. The same scheduling law for $(\beta, \mu)$ is applied and the results are consistent with soft BPGD.

Table \ref{tab1} reports the corresponding results for the soft–hard BPGD encoder (where additional hard code node fixing is applied between message-passing rounds) for three representative block lengths
($N = 100$, $N = 1000$ and $N = 10{,}000$) and rate $R = \tfrac{1}{2}$.
All simulations were performed with 100 message-passing iterations and identical degree distributions.
The columns labeled ``Start point'' and ``End point'' correspond to the initial and final values
of the auxiliary variable~$\xi$ used to compute the softness parameters $(\beta_r,\mu_r)$
according to~\eqref{eq:xi_relation}.
For all code lengths, the results are consistent with soft BPGD where the proposed dynamic schedules (linear and exponential)
consistently reduce average distortion compared with the constant--parameter baseline.
The improvement becomes more pronounced for larger block lengths,
where the exponential schedule achieves the lowest distortion due to its smoother
evolution of~$\xi_r$ across decimation rounds.
These results confirm that 
gradual parameter adaptation enhances convergence and rate-distortion performance
relative to fixed global settings.

\begin{table}[t!]
\setlength{\arraycolsep}{1pt}
\centering
\caption{LDGM code distortion results at rate $R=0.5$
comparing constant, linear, and exponential scheduling.}
\begin{tabularx}{\columnwidth}{cc>{\centering\arraybackslash}p{1cm}ccc}
\hline
\textbf{Length} & \textbf{Iteration} & \textbf{Method} &
\textbf{Distortion} & \textbf{Start point} & \textbf{End point} \\
\hline
100 & 100 & Constant & 0.1561 & 0.050 & 0.050 \\
100 & 100 & Linear & 0.1528 & 0.025 & 0.052 \\
100 & 100 & Exponential & 0.1503 & 0.025 & 0.052 \\
1000 & 100 & Constant & 0.1493 & 0.040 & 0.040 \\
1000 & 100 & Linear & 0.1487 & 0.022 & 0.048 \\
1000 & 100 & Exponential & 0.1476 & 0.022 & 0.048 \\
10000 & 100 & Constant & 0.1463 & 0.030 & 0.030 \\
10000 & 100 & Linear & 0.1426 & 0.012 & 0.032 \\
10000 & 100 & Exponential & 0.1413 & 0.012 & 0.032 \\
\hline
\end{tabularx}
\label{tab1}
\end{table}
\begin{table}[t]
\centering
\caption{Distortion results for soft-hard BPGD for LDGM codes with code nodes with Poisson distribution.}
\label{tab:poisson10k}
\begin{tabular}{cccc}
\toprule
\textbf{ $K$} & \textbf{Constant } & \textbf{Linear } & \textbf{Exponential } \\
\midrule
3 & 0.1389 & 0.1363 & {0.1357} \\
4 & 0.1567 & 0.1496 & {0.1483} \\
5 & 0.1632 & 0.1528 & {0.1512} \\
\bottomrule
\end{tabular}
\end{table}

Table~\ref{tab:poisson10k} shows the average distortion performance of rate $R=1/2$ semi-regular LDGM codes
with a Poisson degree distribution and three fixed generator node degrees $K$.
All simulations were carried out for block length $N=10{,}000$ using
$100$ message--passing iterations.
Three parameter settings were compared: 
 constant softness parameter~$\xi$, 
linearly scheduled~$\xi$, and 
 exponentially scheduled~$\xi$.
Across all generator degrees, both linear and exponential schedules achieved lower distortion than the constant baseline,
demonstrating that dynamic scheduling of $(\beta,\mu)$ improves rate--distortion performance even for uncoupled ensembles.
The exponential schedule consistently yielded the smallest distortion, reflecting its smoother progression of~$\xi_r$
and its ability to maintain stable yet decisive message updates during decimation.
The relative improvement becomes more pronounced as the generator degree~$K$ increases,
indicating that adaptive scheduling compensates for the stronger local constraints in denser graphs.

\section{Conclusions}
We proposed a dynamic parameter scheduling framework for soft and soft-hard BPGD in LDGM-based lossy source coding. By starting with softer parameters and hardening them as decimation proceeds, the encoder aligns its effective temperature with the evolving factor graph and reduces reliance on exhaustive parameter sweeps. Linear and exponential schedules are simple to implement, maintain same-order complexity of the algorithm, and deliver consistent improvements in distortion and convergence across irregular LDGM codes.

\section*{Acknowledgments}
This material is based upon work supported by the National Science Foundation under Grant No. CCF-2145917. 

\bibliographystyle{IEEEtran}

% Paste into your Overleaf .tex file (assumes \usepackage{amsmath,amssymb} in preamble)

\end{document}